\newcommand{\eqn}[1]{\label{#1}}
\newcommand{\eq}[1]{\begin{equation}#1\end{equation}}              
\newcommand{\eqs}[1]{\begin{eqnarray}#1\end{eqnarray}}             
\newcommand{\mx}[2]{\left(\begin{array}{#1}#2\end{array}\right)}   
\newcommand{\bra}[2]{{}_{#2}\big{<}#1\big{|}}			   
\newcommand{\ket}[2]{\big{|}#1\big{>}_{#2} }			   
\def\lb{\nonumber\\}
\newcommand{\refbr}[1]{(\ref{#1})}

\documentstyle[12pt]{article}

\def\d{\delta}

\def\y{\eta}
\def\x{\xi}

\def\l{\lambda}

\def\r{\rho}

\def\s{\sigma}

\def\o{\omega}
\def\G{\Gamma}

\def\/{\over}
\def\*{\partial}

\def\|{\mid}

\def\hatpsi{\hat{\psi}}
\def\barpsi{\bar{\psi}}
\def\hatbarpsi{\hat{\barpsi}}

\def\2{\half}
\def\3{{1\/3}}

\newcommand{\NP}[1]{Nucl. Phys.\ {\bf #1}\ }
\newcommand{\PL}[1]{Phys. Lett.\ {\bf #1}\ }

\newcommand{\NC}[1]{Nuovo Cimento\ {\bf #1}\ }

\newcommand{\IJMP}[1]{Int. J. Mod. Phys. \ {\bf #1} \ }
\newcommand{\half}{\mbox{\scriptsize $ \frac{1}{2}$}}

\include{psfig}
\begin{document}

\newpage
\pagenumbering{arabic}
\begin{flushleft}
G\"{o}teborg\\
ITP 93 -2\\
April 1993
\end{flushleft}
\vspace{1.5cm}
\begin{center}
{\huge Gluing of Branched Surfaces by Sewing of Fermionic String
Vertices}\\[1cm]
{\large Bengt E.W Nilsson\footnote{TFEBN@FY.CHALMERS.SE}\\Per
Sundell\footnote{TFEPSU@FY.CHALMERS.SE}}\\[1cm]
{\sl Institute of Theoretical Physics\\Chalmers University of Technology\\ and
University of G\"{o}teborg\\
S-412 96 G\"{o}teborg, Sweden}
\end{center}
\vspace{1cm}
\begin{abstract}
We glue together two branched spheres by sewing of two Ramond (dual)
two-fermion string vertices and present a rigorous
 analytic derivation of
the closed expression for the four-fermion string vertex. This method treats
all oscillator levels collectively
and the obtained answer verifies that the closed form of the four vertex
previously argued for on the basis of
explicit results restricted to the first two oscillator levels
is the correct one.
\end{abstract}
\newpage

There are by now several different methods available for calculating
correlation functions in string theory and
conformal field theory (CFT). When applied to correlation functions involving
only  untwisted fields, the method
of operator sewing of string (Reggeon) vertices is relatively straightforward
and can be used to produce closed
expressions at arbitrary genus and number of external legs; for a recent review
see \cite{DiV92} and references
therein. The sewn answers are expressable in closed form
in terms of geometrical quantities given in the Schottky representation of the
Riemann
surface in question. However when applied to twisted fields, i.e. fields with
non-integer power expansions in
$z$ or $\bar{z}$ (e.g the Ramond (R) fields of the Neveu-Schwarz-Ramond (NSR)
string), this method has so far
been far less successful. Here the derivation of a closed geometrical form of
the sewn expressions is a step that still remains
to be done.
In fact even for such a fundamental and conceptually simple object as the
vertex for
four external twisted fermions, i.e. the four Ramond vertex, the closed
geometric form is surprisingly difficult to  derive by sewing,
and it is only
recently that the old result for the scattering of massless fermions in the NSR
string \cite{CO,Cor,SW1,CGOS,SW3,BCO}
has been extended (in the matter sector) to the full sewn vertex \cite{ENS2}.
(Some results along these lines have also been obtained recently
 for $\bf{Z}_3$ twisted fermions \cite{EN1}.) This development also made it
clear that the sewn vertex
can be cast into a unique closed geometrical form. That is, it was demonstrated
in \cite{ENS2} that by computing explicitly
 the matrix expressions for a large number of terms in the exponent of the
vertex at oscillator level zero and one,
 one checks easily
that they are all generated by the closed form of the
four Ramond vertex already argued for in \cite{DHMR}.

In CFT the difficulties with twisted fields can be sidestepped by resorting to
various other means of computing
correlations functions,
see for instance \cite{BPZ,ISZ}. As long as one is studying CFT the situation
is in principle quite satisfactory. In string
theory, on the other hand, the method of sewing together fundamental vertices
to produce the different terms in the perturbation
expansion plays a more significant role. This fact becomes particularly clear
in the context of string
field theory when discussed as for instance in \cite{LPP1,LPP2,SZ2}.
Consequently, since sewing involving vertices
containing twisted fields is a poorly developed subject, our understanding of
(open/closed) gauge invariant interacting NSR superstring
field theory (in the twisted sector) is not as detailed and explicit as one
would like. Comparing to the situation for the
bosonic string field theory  where things are rather well under control, a
similar
level of understanding of superstring field theory seems significantly more
difficult to obtain. In particular, following \cite{LPP1,LPP2}
it is possible to show that sewing (at tree level) does produce results which
are connected to the Riemann surface (via the corresponding
propagator) that one naively expects to have generated. This goes under the
name of the Generalized Gluing and Resmoothing Theorem (GGRT)
when the transformations involved in the sewing are general conformal
transformations. As far as the Ramond sector of the
superstring is concerned no results along these lines have yet been obtained.
However, restricting the transformations to projective ones
the explicit results of ref. \cite{ENS2} provides a strong indication for the
validity of the theorem also in the twisted case (at least
in the restricted sense).

In this paper we will improve the situation involving twisted fermion fields by
presenting an analytic
proof of the fact that the two expressions for the four Ramond vertex, i.e. the
one obtained by sewing and the corresponding
closed geometrical form,
discussed in detail in \cite{ENS2}, are equal (both will be given explicitly
below). This proof is valid for all oscillator
levels. The important new feature of our approach is that after the initial
sewing one can recollect all Ramond modes into
 transported Ramond fields and carry through the proof keeping the fields
intact. This then means that the fermion fields
play a far less significant role in the proof and one can focus entirely on the
issue of the equivalence (under
a double integral) of the propagator of the sewn surface and its representation
obtained in the actual sewing.
 We now turn to a brief review of the origin of these two forms of the four
Ramond vertex.

The matrix form of the expression for the four Ramond vertex obtained from
sewing
  is easily derived by inserting a  NS completeness relation
between two Ramond emission vertices \cite{CO}, or equivalently between two
dual Ramond vertices previously derived in \cite{NT}.
Following the latter reference, the four Ramond vertex is obtained by computing
  a correlation function in an auxiliary Hilbert space as follows:
\eq{\hat{W}_{R_1,R_2}(V_1,V_2)=\bra{0}{aux}\hat{W}_{R_1}(V_1)\hat{W}_{R_2}
(V_2)\ket{0}{aux}  \eqn{sewW12}}
where, in terms of the complex fermions defined in \cite{ENS2} (the exact
definition of which will not be relevant
for the rest of this paper), each transported dual Ramond vertex \cite{NT}
reads
\eq{\hat{W}_{R}(V)=\bra{0}{no}\;:exp\oint_{C}dz
\left\{\hat{\bar{\psi}}_{aux}^{V}(z)(\hat{\psi}_{R}+i\hat{\psi}_{no})(z)+
\hat{\psi}_{aux}^{V}(z)(\hat{\bar{\psi}}_{R}+
i\hat{\bar{\psi}}_{no})(z)\right\}:\ket{0}{no} \eqn{dualW}}
Here the contour $C$ encircles the two emission points $V(0)$ and $V(\infty)$
where $V(z)$ is a projective transformation.
Furthermore, the transported fields  are defined by
\eq{\hatpsi^{V}(z)=\sqrt{V'(z)}\;\hatpsi(V(z))  \eqn{defofhatpsiV}}
and $\psi_{aux}$ is a NS (i.e. untwisted) field in the auxiliary Hilbert space
mentioned above. $\psi_{no}$ is an NS normal ordering field;
 performing the correlation function
indicated in the dual vertex gives rise to a form of the vertex which is
explicitly normal ordered also in the external Ramond field
$\psi_{R}$ (the double dots in \refbr{dualW} refer only
to the auxiliary field). This procedure produces a term in the exponent which
is bilinear in the auxiliary field. The vertex
 then becomes
far more tricky to deal with than e.g. an ordinary NS or bosonic (untwisted)
vertex. After having eliminated
 the normal ordering field
one may perform the auxiliary correlation in eq.\refbr{sewW12} by turning it
into an infinite dimensional integral.
For the details of this computation we refer the reader to \cite{ENS2}. This
reference also contains a discussion of the
various choices of projective transformations which make the calculation
tractable. In the present paper we will only
use one choice namely (denoted as choice III in \cite{ENS2})
\eq{V^{-1}_{1}(z)=z+{1\/\l}\;,\;\;\;\;V^{-1}_{2}(z)={{z\/\l}+1 \/ z}
\eqn{V1V2}}
where the hyperelliptic modulus $\l$ satisfies $|\l|<1$.
 This gives
\eqs{\hat{W}_{R_1R_2}(\l)=det(1-M^2)\;
:exp\left(\bar{U}^{T(-)}_{R_2}{1\/1-M^2}U^{(+)}_{R_1}+
U^{T(-)}_{R_2}{1\/1-M^2}\bar{U}^{(+)}_{R_1} \right.
\lb
\left.
+U^{T(-)}_{R_2}{M\/1-M^2}\bar{U}^{(-)}_{R_2}-
\bar{U}^{T(+)}_{R_1}{M\/1-M^2}U^{(+)}_{R_1} \right): \eqn{sewW12matrix}}
where the $\l$ dependent infinite dimensional matrix $M(\l)$ and vectors
$U_{R_i}^{(\pm)}(\l)$ are given by ($r,s$ will in this paper always
refer to positive half integers, and $m,n$ to integers):
\eq{M_{rs}(\l)=\oint_{0}dz\;\oint_{0}dw\;
z^{-r-{1\/2}}w^{-s-{1\/2}}{1-\sqrt{{w+{1\/\l}\/z+{1\/\l}}}\/z-w}
={r\/r+s}\mx{l}{-{1\/2}\\r-{1\/2}}
\mx{l}{-{1\/2}\\s-{1\/2}}\l^{r+s}\eqn{defofM}}
\eqs{\bar{U}^{(+)}_{R_1,r}(\l)=\bar{U}^{(+)}_{rm}
\hatbarpsi^{R_{1}}_{m}=\sqrt{2}v^{(m)}_r\;\hat{\bar{\psi}}^{R_1}_m\;,
\;\;\;\;U^{(+)}_{R_1,r}=U^{(+)}_{rm}\hatpsi^{R_{1}}_{m}=
\sqrt{2}v^{(m)}_r\;\hat{\psi}^{R_1}_m\;, \lb
\bar{U}^{(-)}_{R_2,r}=\bar{U}^{(-)}_{rm}\hatbarpsi^{R_{2}}_{m}=
-i\sqrt{2}v^{(m)}_r\;\hat{\bar{\psi}}^{R_2}_m\;,
\;\;\;\;U^{(-)}_{R_2,r}=U^{(-)}_{rm}\hatpsi^{R_{2}}_{m}=
-i\sqrt{2}v^{(m)}_r\;\hat{\psi}^{R_2}_m\;,\eqn{detofU}}
where
\eq{v^{(n)}_{r}={1\/\sqrt{2}}\mx{l}{-n-\2\\r-\2}\l^{r+n} \eqn{defofvn}}

 One of the main results of ref. \cite{ENS2} was to show that there exists an
algorithm
for calculating, for any integers $m,n$, the quantity $(U^T{1\/1-M^2}U)_{mn}$
appearing in the exponent of the vertex multiplying
two different Ramond fields, as well as  $(U^T{M\/1-M^2}U)_{mn}$ which multiply
two identical Ramond fields. It was also found there
that the results of these calculations could be obtained form a closed form of
the whole vertex, or, more precisely, that
the different terms obtained in the sewing process relate to one and the same
propagator of the sewn surface. Hence, in the
language of \cite{LPP1,LPP2} this provides a first step towards proving a
restricted form of the GGRT for twisted fermions although
the results were not discussed in these terms in \cite{ENS2}.

\begin{figure}
\centerline{
\psfig{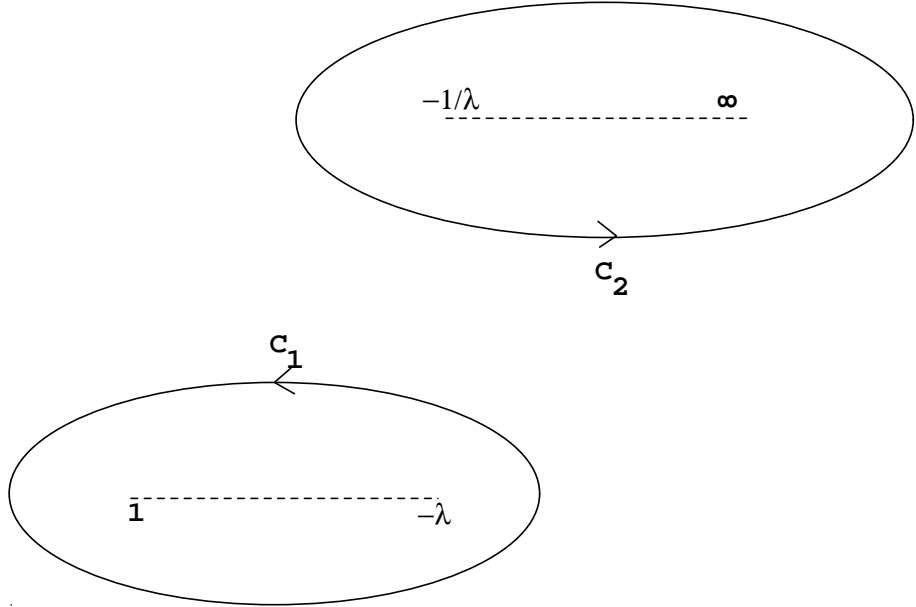}
}
\caption{The integration contours $C_{i}$.}
\label{figurett}
\end{figure}

 Before presenting the closed form of the vertex expressed in terms of the
above choice of projective transformations we will give it
in the
following more general version ($i,j=1,2$)
\eq{\hat{W}_{R_1R_2}(V_1,V_2)=\left({V^{-1}_{2}
(z_{1}^{(1)})\/V^{-1}_{2}(z_{1}^{(2)})}\right)^{{1\/4}}
:exp\left(\sum_{i,j}\oint_{C_{i}}dz\oint_{C_{j}}dw\;
\hatpsi_{i}^{V^{-1}}(z)G(V_{1},V_{2};z,w)\hatbarpsi_{j}^{V^{-1}}(w)\right):
\eqn{theclosedvertex}}
where the normal ordering refers to the oscillators of both complex Ramond
fields, and
 the propagator $G(V_{1},V_{2};z,w)$ of the produced surface is given by
\eq{G(V_{1},V_{2};z,w)={1\/z-w}\sqrt{{V^{-1}_{1}(w)
V^{-1}_{2}(w)\/V^{-1}_{1}(z)V^{-1}_{2}(z)}}\eqn{gv1v2}}
Furthermore, as shown in figure \refbr{figurett}, the closed contours $C_{i}$
in
\refbr{theclosedvertex} encircle the cuts in the twisted (Ramond) fermion
fields $\hatpsi_{i}^{V^{-1}}(z)$ defined by
\eq{\hatpsi_{i}^{V^{-1}}(z)=\sqrt{(V^{-1}_{i})'(z)}
\hatpsi_{R_{i}}(V^{-1}_{i}(z))\eqn{defofhatpsii}}
where $V_{i}$ are the projective transformations by means of which the branch
points of the twisted fields
have been transported from $0,\infty$ to the emission points
$z_{1,2}^{(i)}=V_{i}(0),V_{i}(\infty)$, that is
$V_{i}^{-1}(z)=k_{i}{z-z_{1}^{(i)}\/z-z_{2}^{(i)}}$. Note that the integrals in
the exponent of \refbr{theclosedvertex}
are easily performed since the branch cuts in the Ramond fields cancel  against
similar cuts in the propagator
$G(V_1,V_2;z,w)$ and the contours actually encircle  only poles at
$z^{(i)}_{1,2}$.

As mentioned in \cite{ENS2}, the above closed form of the vertex should be
valid for any choice of projective transformations
$V_1$ and $V_2$ but results to this effect exist so far only for the four
particular choices discussed in \cite{ENS2}.
Although the analytic proof given below will be presented  only for the one
choice of $V$'s given above it can easily be
 repeated for any of the other three choices of ref. \cite{ENS2}. The ultimate
goal must be to find a proof
independent of any particular choice of $V$'s which very likely would be of
great help in extending these results
to six or more external Ramond legs. We now leave the general discussion and
return to the particular choice of
 $V$'s given above in \refbr{V1V2}.

The proof that the two forms of the vertex described above are equivalent for
the projective transformations in eq.\refbr{V1V2}
 involves, besides
 showing the equivalence
of the prefactors which is trivial \cite{ENS2}, proving that the exponents are
analytically identical. In the following we will
refer to the exponent coming from sewing of two dual Ramond vertices as the
LHS, and to the exponent appearing in the
 closed form as the RHS. Hence we want to prove that
LHS=RHS or in other words that
\eq{{ \oint_{C_{i}} dz \oint_{C_{j}} dw
\;\sum_{r,s}\hatpsi_{i}^{\l}(z)z^{(-1)^{i}r-{1\/2}} B^{ij}_{rs}
(\l)w^{(-1)^{j}s-{1\/2}}
\hatbarpsi_{j}^{\l}(w)=\oint_{C_{i}}dz\oint_{C_{j}}dw\;
\hatpsi_{i}^{\l}(z)G(\l;z,w)\hatbarpsi_{j}^{\l}(w)}\eqn{lhsequalsrhs}}
Note that on the LHS we have reintroduced the Ramond fields $\psi_i^{\l}(z)$
(we use the superindex $\l$ instead of $V_{i}^{-1}$ when
refering to the projective transformations in \refbr{V1V2}) by reconstructing
them from the matrices
$U_{rm}^{(\pm)}(\l)$ and the respective Ramond oscillators $\psi^{R_{i}}_m$ and
similarly for the barred quantities.
Furthermore, as can be read off from \refbr{gv1v2}, \refbr{V1V2} and
\refbr{sewW12matrix}:
\eqs{G(\l;z,w)&=&{1\/z-w}\sqrt{(1+\l w)(w+\l)z\/(1+\l
z)(z+\l)w}\eqn{defofglambda}\\
B^{11}(\l)&=&-B^{22}(\l)={M(\l)\/1-M^{2}(\l)}\\
B^{12}(\l)&=&-B^{21}(\l)={1\/1-M^{2}(\l)}\eqn{gabdboflambda}}
where the matrix $M(\l)$ is given in \refbr{defofM}. The equality
\refbr{lhsequalsrhs} simply means that $B^{ij}(\l)$ is the matrix of
Taylor coefficients of the Taylor expansion of the propagator $G(\l;z,w)$ for
$z\in C_{i}$ and $w\in C_{j}$.
Note that on the LHS the explicit powers of $z$ and $w$ depend on $i,j=1,2$.
 This is just a consequence of the radial ordering of
the two dual vertices in the original auxiliary correlation function that led
to the four vertex. However, also the quantities
$B_{rs}$ depend on which fermionic fields they multiply. By combining these
facts one finds that this dependence on the external
fields disappears, as can be seen explicitly
on the RHS. The occurance of this phenomenon is of course no surprise since the
propagator $G$ should
not depend on which external Ramond fields it connects.
 In a geometrical language, proving this equation means that we have performed
an explicit fermionic gluing of two
branched spheres (with two branch points), defined by the two-point functions
appearing in the two dual Ramond vertices,
into one genus one hyperelliptic surface (sphere with four branch points),
defined by the two-point propagator $G$ appearing in the
Ramond four-vertex.

We begin the proof of the equivalence between the LHS and the RHS by deriving
integral representations
for the matrices $B^{ij}(\l)$ and the propagator $G(\l;z,w)$. These are
obtained by acting with $\int d\l\*_{\l}$:
\eqs{B^{ij}(\l)&=&B^{ij}(0)+\int_{0}^{\l}{dx\/x}x\*_{x}B^{ij}(x)
\eqn{intrepofbijs}\\
G(\l;z,w)&=&G(0;z,w)+\int_{0}^{\l}{dx\/x}x\*_{x}G(x;z,w)\eqn{intrepofg}}
(For the quantities $B^{ij}$, this trick was used for twisted scalars in
\cite{DGM2}.) These integral
representations will play a crucial role in the following. To understand the
reason for this we will first
discuss the consequences of rewriting $B^{ij}$ this way. Later we will see that
rewriting $G$ leads to similar
results. This will facilitate the comparison of the LHS and the RHS, and help
us identify a final subtlety that
will be discussed later.

We start by observing that
\eqs{B^{11}&=&-B^{22}=\half((1-M)^{-1}-(1+M)^{-1})\lb
B^{12}&=&-B^{21}=\half((1-M)^{-1}+(1+M)^{-1}) \eqn{defofB}}
and  that the logarithmic derivative of $M$ has rank zero, or more
precisely
\eq{\l\*_{\l}M(\l)=2Rvv^{T}\eqn{lambdaderivativeofm}}
where the components of the vector $v$ are
\eq{v_{r}(\l)=v_{r}^{(0)}(\l)={1\/\sqrt{2}}\mx{l}{-{1\/2}\\r-{1\/2}}\l^{r}\;\;
,\eqn{defofv}}
$R_{rs}=r\d_{rs}$ and $T$ denotes transposition.
Using \refbr{lambdaderivativeofm} and the following equation derived in
\cite{ENS2},
\eqs{&&(1\pm M^{T})^{-1}v=2g^{\mp}(1\mp M)^{-1}v\lb
&&g(\l)={1+\l\/1-\l}\;\;\; ,\nonumber}
we find
\eqs{\l\*_{\l}(1-M)^{-1}&=&2gR\x\x^{T}\lb
\l\*_{\l}(1+M)^{-1}&=&-2g^{-1}R\y\y^{T}\nonumber}
where we have adopted the definitions $\x\equiv (1+M)^{-1}v$ and $\y\equiv
(1-M)^{-1}v$.
Inserting these results into \refbr{defofB} gives
\eqs{B^{11}(\l)&=&-B^{22}(\l)=\int_{0}^{\l}{dx\/x}
(g(x)R\x(x)\x^{T}(x)+g(x)^{-1}R\y(x)\y^{T}(x))\eqn{intrepofa11}\\
B^{12}(\l)&=&-B^{21}(\l)=1+\int_{0}^{\l}{dx\/x}
(g(x)R\x(x)\x^{T}(x)-g(x)^{-1}R\y(x)\y^{T}(x))\eqn{intrepofa12}}

\begin{figure}
\centerline{
\psfig{figure=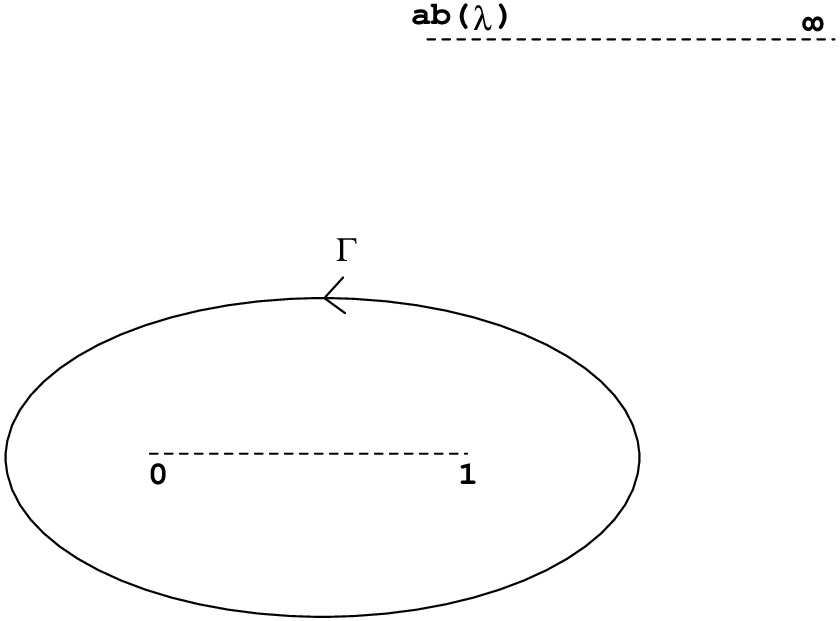,height=8cm,width=12cm}
}
\caption{The contour $\G$ used to define $\x_{r}$ and $\y_{r}$.}
\label{figurtvaa}
\end{figure}

The next piece of information that will be of importance is that one may derive
explicit solutions\footnote{These solutions are
hypergeometric $_{2}F_{1}$ functions written here in terms of their integral
representations.} for the vectors
$\x\equiv (1+M)^{-1}v$ and $\y\equiv (1-M)^{-1}v$:
\eqs{\x_{r}(\l)&=&{1\/\sqrt{2}}\oint_{\G}{dt\/t}\o(\l;t)^{\pm
r}={ab-1\/\sqrt{2}}\oint_{\G}{dt\/(t-1)(t-ab)}\o(\l;t)^{\pm r}\eqn{defofxs}\\
\y_{r}(\l)&=&\pm{ab\/\sqrt{2}}\oint_{\G}{dt\/t(t-ab)}\o(\l;t)^{\pm
r}=\pm{1\/\sqrt{2}}\oint_{\G}{dt\/t-1}\o(\l;t)^{\pm r}\eqn{defofys}}
where
\eq{\o(\l;t)={t-ab\/t(1-t)}\eqn{defofomega}}
and
\eq{a(\l)={1+\l\/2}\;\;\;\;\;\;,\;\;\;\; b(\l)={1+\l\/2\l}\eqn{defofaandb}}
The contour $\G$ used in \refbr{defofxs} and \refbr{defofys} encircles the
branch cut between $0$ and $1$, as shown in figure
\refbr{figurtvaa}. The first integral form for $\x_r(\l)$ appearing in
\refbr{defofxs} was derived in \cite{CGOS} while formulas  related
to $\y_r(\l)$ can be found in \cite{CGOS,SW3}. The four different forms of
integral representations in \refbr{defofxs} and \refbr{defofys}
are related by total derivative terms or
 by the reparametrization  $t\rightarrow {ab\/t}$.

 In $B^{ij}_{rs}$ any combinations of $\x$'s and $\y$'s may be
used but the computations will in this paper  only be carried out in detail for
one particular choice. However, for any choice we see
that since the only $r,s$ dependence of the intergral representations of
$\x_{r},\y_{r}$ and $\x_{s},\y_{s}$ comes from
$\o(x;t)^{\pm r}$, say, and $\o(x;u)^{\pm s}$, say, the sums in
the LHS are trivial. By flipping the contours $C_{i}$ and $C_{j}$
to the poles of the sums at $z=\o(x;t)$ and $w=\o(x;u)$ we realize that the
arguments of the fermi-fields $\hatpsi_{i}$ and
$\hatbarpsi_{j}$ will be the functions $\o(x;t)$ and $\o(x;u)$, respectively.
We also see from \refbr{intrepofa11} and
\refbr{intrepofa12} that in each term inside the integration over $x$ the
dependence on $t$ and $u$ will be completely factorized.
These two facts will guide us when we rewrite the integral representation of
$G$. However, for a completely rigorous treatment
of the LHS we must ensure convergence in the sums
by picking appropriate contours in the integral representations of $\x$ and
$\y$. We will return to this question after having
discussed the integral representation of $G$.

Let us now study the effect of inserting the integral representation
\refbr{intrepofg} of $G$ into the RHS.
Using \refbr{defofglambda} in \refbr{intrepofg} one finds
\eqs{G(\l;z,w)&=&{1\/z-w}+{1\/2}\int_{0}^{\l}dx\left[-(1+xw)^{-{1\/2}}
(1+xz)^{-{3\/2}}(w+x)^{{1\/2}}(z+x)^{-{1\/2}}+\right.\lb
&&\left.+(1+xw)^{{1\/2}}(1+xz)^{-{1\/2}}(w+x)^{-{1\/2}}
(z+x)^{-{3\/2}}\right]\big{(}{z\/w}\big{)}^{\2}\eqn{intrepofgg}}
Note that the $z$ and $w$ dependence have factorized in each term of the
integrand, just like it did
for $t$ and $u$ in the LHS as explained above. Furthermore, we know that
the argument of the fermion fields in the LHS will be the function
$\o(x;\cdot)$. From the following equations we will
understand why this function suggests itself as natural variables also on the
RHS:
\eqs{1+x\o(x;t)&=&-{(t-a(x))^{2}\/t(1-t)}\lb
\o(x;t)+x=&=&-x{(t-b(x))^{2}\/t(1-t)}\;\;\;,}
Thus, by the changes of variable
\eqs{&&z=\o(x;t)\\
&&w=\o(x;u)\eqn{proposedchangeofvariable}}
all the cuts in the RHS will be gathered into one single factor,
namely $\sqrt{{\o(x;t)\/ \o(x;u)}}$.

\begin{figure}
\centerline{
\psfig{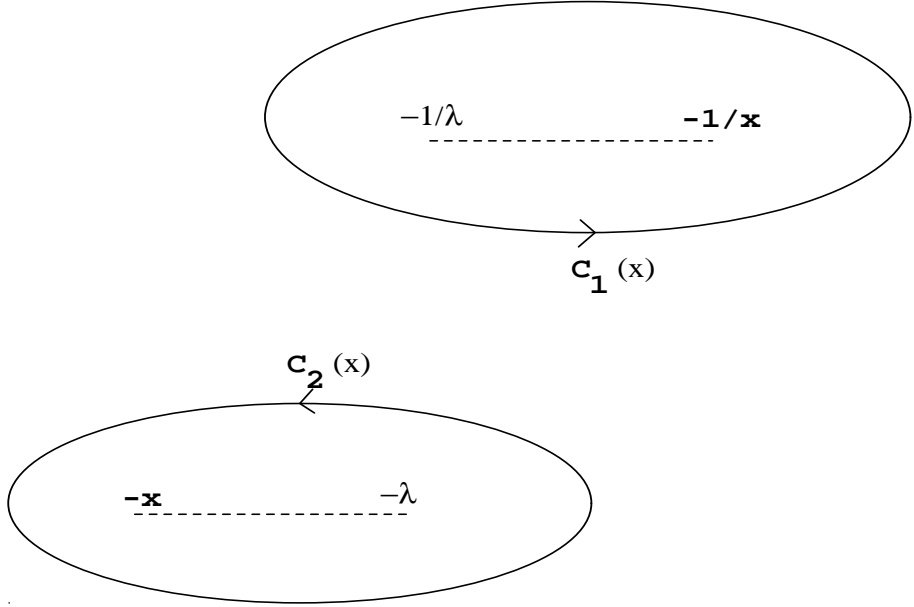}
}
\caption{The $x$-dependent branch cuts and contours in eq. (34). }
\label{figurtre}
\end{figure}

However, before we do this change of variables, we must first justify the
commutation of the order of integration in the RHS from
$\oint_{C_{i}}dz\oint_{C_{j}}dw\int_{0}^{\l}dx$ to
$\int_{0}^{\l}dx\oint_{C_{i}(x)}dz\oint_{C_{j}(x)}dw$. This manouvre induces a
global $x$-dependence in the contours in $z$- and
$w$-planes, as indicated above, since in the latter case there is no longer any
cancellation of branch cuts inside $C_{i}$ in
the integrand. E.g., instead of cancelling each
other, the ($\l$-dependent) branch cut from $\infty$ to $-{1\/\l}$ in
$\hatpsi^{\l}_{1}(z)$ and the corresponding ($x$-dependent)
branch cut from $-{1\/x}$ and $\infty$ in the integral representation
\refbr{intrepofgg} of $G$ now merge into a new branch
cut from $-{1\/\l}$ to $-{1\/x}$ which has to be encircled by $C_{1}(x)$. In
figure \refbr{figurtre} we show both $C_{1}(x)$
and $C_{2}(x)$.

\begin{figure}
\centerline{
\psfig{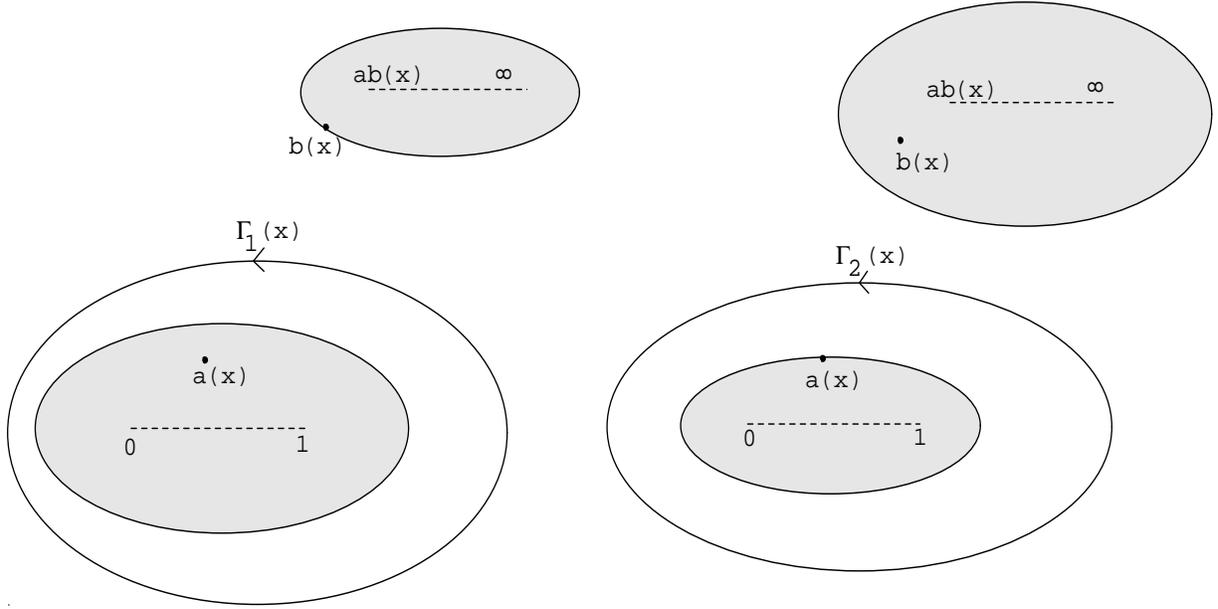}
}
\caption{Two complex planes containing the proper contour $\G_{i}(x)$. The
shaded regions are forbidden by the conditions in
eq. (31).}
\label{figurfyra}
\end{figure}

We will from now on assume that $|x|<|\l|$ in $\int_{0}^{\l}dx$. Thus in the
proposed change of variables, namely
$z=\o(x;t)$ and $w=\o(x;u)$,
we must choose contours $\G_{i}(x)$ in the $t$-plane (and $\G_{j}(x)$ in the
$u$-plane) such that
\eqs{|x|&<&|\o(x;t)|<|{1\/\l}|\;\;\;\;\;\;\;\;{\rm
for\;all}\;\;t\in\G_{1}(x)\lb
|\l|&<&|\o(x;t)|<|{1\/x}|\;\;\;\;\;\;\;\;{\rm for\;all}\;\;t\in\G_{2}(x)}
Since $|x|<|\l|$, and since
\eqs{&&\o(x;0)=\o(x;1)=\infty\lb
&&\o(x;a(x))=-{1\/x}\;\;\;,\;\;\;\o(x;b(x))=-x\lb
&&\o(x;a(x)b(x))=\o(x;\infty)=0\eqn{someformulae}}
this means that proper contours $\G_{i}(x)$ has to be chosen as in figure
\refbr{figurfyra}.

Moreover, as $t$ encircles $\G_{i}(x)$ once $\o(x;t)$ encircles $C_{i}$ twice
in the positive direction for $i=1$ but in
the negative direction for $i=2$. Thus, using also
\eq{\*_{t}\o(x;t)={(t-a(x))(t-b(x))\/t^{2}(t-1)^{2}}}
we get (here we also show the intermediate steps discussed above)
\eqs{&&\oint_{C_{i}}dz\oint_{C_{j}}dw\;
\hatpsi^{\l}_{i}(z)G(\l;z,w)\hatbarpsi^{\l}_{j}(w)=\lb
&&=\oint_{C_{i}}dz\oint_{C_{j}}dw\;
\hatpsi^{\l}_{i}(z){1\/z-w}\hatbarpsi^{\l}_{j}(w)+\lb
&&+{1\/2}\int_{0}^{\l}dx\oint_{C_{i}(x)}dz\oint_{C_{j}(x)}dw\hatpsi^{\l}_{i}(z)
\left[-(1+xw)^{-{1\/2}}(1+xz)^{-{3\/2}}(w+x)^{{1\/2}}(z+x)^{-{1\/2}}+\right.\lb
&&\left.+(1+xw)^{{1\/2}}(1+xz)^{-{1\/2}}(w+x)^{-{1\/2}}
(z+x)^{-{3\/2}}\right]\big{(}{z\/w}\big{)}^{\2}
\hatbarpsi^{\l}_{j}(w)=\eqn{intermediate}\\
&&=\oint_{C_{i}}dz\oint_{C_{j}}dw\;
\hatpsi^{\l}_{i}(z){1\/z-w}\hatbarpsi^{\l}_{j}(w)+\lb
&&+{1\/2}\int_{0}^{\l}dx{1\/4}(-1)^{i+j}
\oint_{\G_{i}(x)}dt{(t-a)(t-b)\/t^{2}(t-1)^{2}} \hatpsi^{\l}_{i}(\o(x;t))
\oint_{\G_{j}(x)}du{(u-a)(u-b)\/u^{2}(u-1)^{2}}\hatbarpsi^{\l}_{j}(\o(x;u))\lb
&&\left[-{u-b\/u-a}{t^{2}(t-1)^{2}\/(t-a)^{3}(t-b)}+
{1\/x^{2}}{u-a\/u-b}{t^{2}(t-1)^{2}\/(t-a)(t-b)^{3}}\right]
\left({\o(x;t)\/\o(x;u)}\right)^{\2}=\lb
&&=\oint_{C_{i}}dz\oint_{C_{j}}dw\;\hatpsi^{\l}_{i}(z){1\/z-w}
\hatbarpsi^{\l}_{j}(w)+\lb
&&-{1\/2}(-1)^{i+j}\int_{0}^{\l}{dx\/x}{ab\/g(x)}\oint_{\G_{i}(x)}
{dt\/(t-a)^{2}(t-b)^{2}}\sqrt{\o(x;t)}\hatpsi^{\l}_{i}(\o(x;t))\lb
&&\oint_{\G_{j}(x)}{du\/u(u-1)(u-ab)}\sqrt{\o(x;u)}\hatbarpsi^{\l}_{j}(\o(x;u))
\left[t^{2}u^{2}-t^{2}u-tu^{2}+ab(u+t)-(ab)^{2}\right]\eqn{rhs}}
where one should note that all $a$'s and $b$'s depend on $x$. This will be our
final result for the RHS. When we now
return to the LHS we will discover that it can be brought to an almost
identical form. At the end it will shown that this discrepancy is a total
derivative and thus vanishes under the integral.

\begin{figure}
\centerline{
\psfig{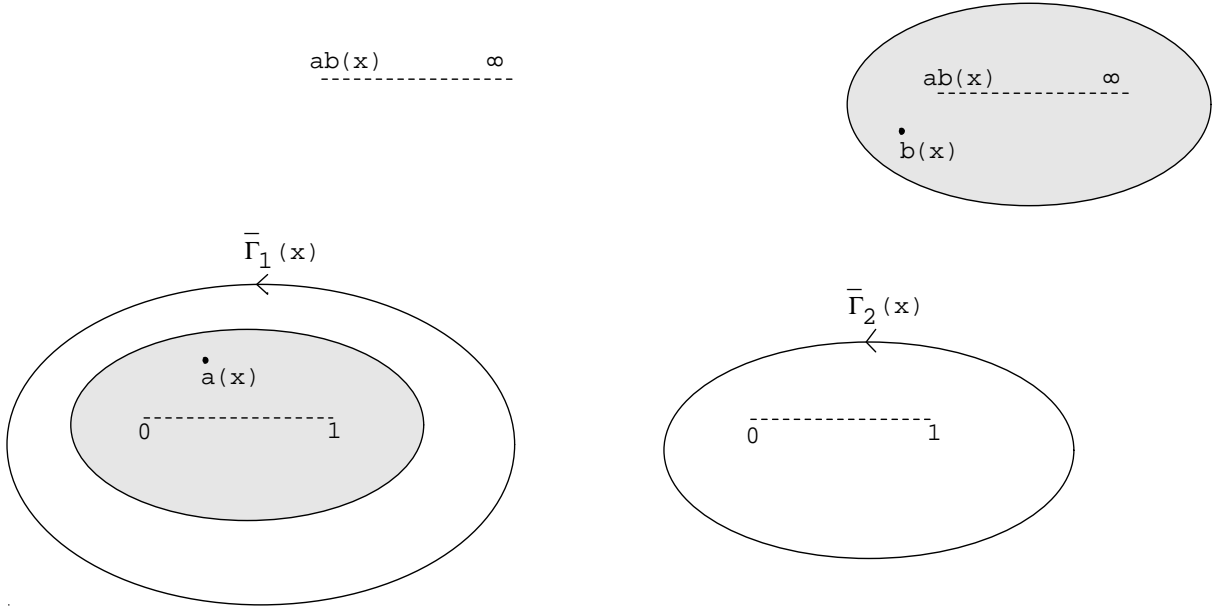}
}
\caption{Proper contours $\bar{\G}_{i}(x)$ to be used in the LHS. The shaded
regions are forbidden by the conditions for
convergence of the sums in eq. (36).}
\label{figurfem}
\end{figure}

Returning to the LHS we insert the integral representations
\refbr{intrepofa11} and \refbr{intrepofa12} of the $B^{ij}$-matrices:
\eqs{&&\oint_{C_{i}}dz\oint_{C_{j}}dw\hatpsi^{\l}_{i}(z)z^{(-1)^{i}r-\2}
B^{ij}_{rs}w^{(-1)^{j}s-\2}\hatbarpsi^{\l}_{j}(w)=\lb
&&=(-1)^{i+1}\oint_{C_{i}}dz\oint_{C_{j}}dw\hatpsi^{\l}_{i}
(z)z^{(-1)^{i}r-\2}\left[(1-\d_{ij})\d_{rs}+\right.\lb
&&\left.+\int_{0}^{\l}{dx\/x}(g(x)r\x_{r}(x)\x_{s}(x)+(-1)^{i+j}
g(x)^{-1}r\y_{r}(x)\y_{s}(x))\right]w^{(-1)^{j}s-\2}\hatbarpsi^{\l}_{j}(w)}
To be specific, let us pick the following integral representations for
$\x_{r,s}(x)$ and $\y_{r,s}(x)$:
\eqs{\x_{r}(x)&=&{1\/\sqrt{2}}\oint_{\bar{\G}_{i}(x)}{dt\/t}\o(x;t)^{-(-1)^{i}
r}\lb
\x_{s}(x)&=&{ab-1\/\sqrt{2}}\oint_{\bar{\G}_{j}(x)}
{du\/(u-1)(u-ab)}\o(x;u)^{-(-1)^{j}r}\lb
\y_{r}(x)&=&(-1)^{i+1}{ab\/\sqrt{2}}\oint_{\bar{\G}_{i}(x)}
{dt\/t(t-ab)}\o(x;t)^{-(-1)^{i}r}\lb
\y_{s}(x)&=&(-1)^{j+1}{1\/\sqrt{2}}\oint_{\bar{\G}_{j}(x)}
{du\/u-1}\o(x;u)^{-(-1)^{j}r}\eqn{choiceofxsandys}}
To ensure convergence of the sums over $r,s$ the contours $\bar{\G}_{i}(x)$ are
such that $\big{|}z/\o(x;t)\big{|}^{(-1)^{i}}<1$
for all $z\in C_{i}$ and all
$t\in \bar{\G}_{i}(x)$, and analogously with $z\rightarrow w$, $i\rightarrow j$
and $t\rightarrow u$.
Here we assume that the contours $C_{i}$ have been collapsed on the branch cuts
that
they are to encircle in such a way that $|z|>{1\/|\l|}$ for
all $z\in C_{1}$ and $|z|<|\l|$ for all $z\in C_{2}$, and analogously for
$z\rightarrow w$.
Again using \refbr{someformulae}, we see that $\bar{\G}_{i}(x)$ have to
be drawn as in \refbr{figurfem}. The sums in the LHS are now well-defined, and
we get
\eqs{&&\oint_{C_{i}}dz\oint_{C_{j}}dw\hatpsi^{\l}_{i}(z)
{1\/z-w}\hatbarpsi^{\l}_{j}(w)\;(1-\d_{ij})+\lb
&&+\2
(-1)^{i+j}\oint_{C_{i}}dz\oint_{C_{j}}dw\hatpsi^{\l}_{i}
(z)\left[\int_{0}^{\l}{dx\/x}\oint_{\bar{\G}_{i}(x)}dt\sqrt{\o(x;t)}
\oint_{\bar{\G}_{j}(x)}du\sqrt{\o(x;u)}\right.\eqn{afterusedintrep}\\
&&\left.({\o(x;t)\/(z-\o(x;t))^{2}}+\2
{1\/z-\o(x;t)})({g(x)(ab-1)\/t(u-1)(u-ab)}+
{g(x)^{-1}(ab)^{2}\/t(t-ab)u(u-ab)}){1\/w-\o(x;u)}\right]
\hatbarpsi^{\l}_{j}(w)\nonumber}

The first term in \refbr{afterusedintrep} agrees with the closed result in
\refbr{rhs} due to fact that
when $i=j$ the first term in \refbr{rhs} vanishes:
\eqs{&&\oint_{C_{i}}dz\oint_{C_{i}}dw\hatpsi^{\l}_{i}(z)
{1\/z-w}\hatbarpsi^{\l}_{i}(w)=
\oint_{C_{i}}dz\big{(}-\oint_{z}\big{)}dw\hatpsi^{\l}_{i}
(z){1\/z-w}\hatbarpsi^{\l}_{i}(w)=\lb
&&=\oint_{C_{i}}dz\hatpsi^{\l}_{i}(z)\hatbarpsi^{\l}_{i}(z)=0\;\;,}
since in the last expression $C_{i}$ encircles all the singularities of the
integrand\footnote{Note that the pole at infinity of
${1\/z-w}$ does not give a contribution to the integral over $w$.}

In the second term in \refbr{afterusedintrep}
the contours $C_{i}$ and $C_{j}$ (which are collapsed on the respective branch
cuts) and $\bar{\G}_{i}(x)$ and $\bar{\G}_{j}(x)$
are precisely such that change of the order of integration between
$\oint_{C_{i}}dz\oint_{C_{j}}dw$ and
$\int_{0}^{\l}dx\oint_{\bar{\G}_{i}(x)}dt\oint_{\bar{\G}_{j}(x)}du$ is allowed
and
such that $C_{i}$ and $C_{j}$
can be flipped over to the poles at $z=\o(x;t)$ and $w=\o(x;u)$. Using also the
identity $(ab-1)g={ab\/g}$ the second term becomes
\eqs{&&\2(-1)^{i+j}\int_{0}^{\l}{dx\/x}{ab\/g(x)}
\oint_{\bar{\G}_{i}(x)}dt\sqrt{\o(x;t)} \oint_{\bar{\G}_{j}(x)}du\sqrt{\o(x;u)}
({\o(x;t)\/\*_{t}\o(x;t)}\*_{t}\hatpsi^{\l}_{i}(\o(x;t))+
\2\hatpsi^{\l}_{i}(\o(x;t)))\lb
&&({1\/t(t-ab)(u-ab)}+{ab\/t(t-ab)u(u-ab)})
\hatbarpsi^{\l}_{j}(\o(x;u))\eqn{afterdonepoles}}
We next remove the $t$-derivative on $\hatpsi^{\l}_{i}$ by integrating by
parts. After some simplifications we arrive at
\eqs{&&-\2(-1)^{i+j}\int_{0}^{\l}{dx\/x}{ab\/g(x)}
\oint_{\bar{\G}_{i}(x)}{dt\/(t-a)^{2}(t-b)^{2}}
\sqrt{\o(x;t)}\hatpsi^{\l}_{i}(\o(x;t))
\eqn{afterpartint}\\
&&\oint_{\bar{\G}_{j}(x)}{du\/u(u-1)(u-ab)}
\sqrt{\o(x;u)}\hatbarpsi^{\l}_{j}(\o(x;u))
\left[u(ab-1)(t^{2}-ab)+(u-1)ab(t^{2}-2t+ab)\right]
\nonumber}

Before comparing \refbr{afterpartint} to the second term in \refbr{rhs} there
is a slight subtlety in \refbr{afterpartint}
which needs to be explained, namely the fact that there is no prescription
of how the contours $\bar{\G}_{i}(x)$ are to encircle the poles at $t=a$ and
$t=b$. However, the reason for this is simply that
these poles have zero residues. These poles stem from the factor
$(\*_{t}\o(x;t))^{-1}$ which was inserted in
\refbr{afterdonepoles} in order to compensate for the
inner derivative of $\*_{t}\hatpsi^{\l}_{i}(\o(x;t))$. We may write the
$t$-integral in \refbr{afterdonepoles} as
\eqs{&&\oint_{\bar{\G}_{i}(x)}dt{f(t)\/(t-a)(t-b)}
\*_{t}\hatpsi^{\l}_{i}(\o(x;t))=
-\oint_{\bar{\G}_{i}(x)}dt\*_{t}({f(t)\/(t-a)(t-b)})
\hatpsi^{\l}_{i}(\o(x;t))=\lb
&&=\oint_{\bar{\G}_{i}(x)}dt\big{(}{f(t)\/(t-a)^{2}
(t-b)}+{f(t)\/(t-a)(t-b)^{2}}-{f'(t)\/(t-a)(t-b)}
\big{)}\hatpsi^{\l}_{i}(\o(x;t))}
where $f$ is analytic at $t=a$ and $t=b$. Thus the residue at e.g. $t=b$ is
given by
\eqs{&&{f(b)\/(b-a)^{2}}\hatpsi^{\l}_{i}(\o(x;b))+
\*_{t}\left.\big{(}{f(t)\hatpsi^{\l}_{i}(\o(x;t))\/t-a} \big{)}\right|_{t=b}-
{f'(b)\/b-a}\hatpsi^{\l}_{i}(\o(x;b))=\lb
&&={f(b)\/b-a}{\hatpsi'_{i}}{}^{\l}(\o(x;b))\*_{t}
\left.\o(x;t)\right|_{t=b}=0\nonumber}
Hence the contours $\bar{\G}_{i}(x)$ in \refbr{afterpartint} and $\G_{i}(x)$ in
\refbr{rhs} are equivalent. The same
holds for $\bar{\G}_{j}(x)$ and $\G_{j}(x)$.

Since the contours of integration in the LHS \refbr{afterpartint} (see
fig.\refbr{figurfem}) and the RHS \refbr{rhs}
(see fig.\refbr{figurfyra}) are
the same, we can now compare the integrand of these two expressions. We then
find that
they differ only in the square bracket and that this difference, namely
\eqs{&&\left[t^2u^2-(t^2u+tu^2)+ab(u+t)-(ab)^2\right]-
\left[u(ab-1)(t^2-ab)+(u-1)ab(t^2-2t+ab)\right]=\lb
&&=t(t-1)(u-a)(u-b)\;,}
gives a vanishing  contribution to the integrals over $u$.
This can be seen as follows. The contour $\G_{j}(x)$ in \refbr{rhs} (which is
now forbidden only in the shaded region in fig.
\refbr{figurtre} which contains
the branch cut in $\hatbarpsi^{\l}_{j}(\o(x;u))$) is precisely such that
$\hatbarpsi^{\l}_{j}(\o(x;u))$ may
be expanded in an analytic Taylor series in $\o(x;u)$ for all $u\in
\G_{j}(x)$\footnote{Note that the Taylor expansion
of $\hatbarpsi^{\l}_{j}(\o(x;u))$ in $\o(x;u)$ requires an interchange
of the order of integration and summation. This manipulation is well-defined,
for instance if the four-fermion vertex is saturated with
external Fock-space states.}. Moreover, for positive integers $n$ (in fact we
could take $n\in{\bf Z}$):
\eqs{&&\oint_{\G_{j}(x)}{du\/u(1-u)(u-ab)}
\sqrt{\o(x;u)}\o(x;u)^{n}(u-a)(u-b)=\lb
&&=\oint_{\G_{j}(x)}\*_{u}\o(x;u)
\o(x;u)^{n-{1\/2}}=2\oint_{\infty}dz\;z^{2n}=0}
Thus:
\eq{\oint_{\G_{j}(x)}{du\/u(1-u)(u-ab)}
\sqrt{\o(x;u)}\hatbarpsi^{\l}_{j}(\o(x;u))(u-a)(u-b)=0}
This completes the proof that the two sides of eq.\refbr{lhsequalsrhs} are
equal.

Finally we comment that there is nothing particular about the above choice
\refbr{choiceofxsandys} of integral representations
for $\x_{r,s}$ and
$\y_{r,s}$. Picking another combination of $\x_{r,s}$ and $\y_{r,s}$ from
\refbr{defofxs} and \refbr{defofys} would
only modify the
$t$ and $x$ dependence in front of the total $u$-derivative that arises when
computing
 the difference between the RHS and the LHS, and would therefore
not affect the (unique) propagator $G(\l;z,w)$.

There are several directions in which one would like to extend the techniques
and results of this paper.
 We have already mentioned in the introduction the importance of proving the
full GGRT (first discussed in the
context of the bosonic string in \cite{LPP1,LPP2}) for twisted fermions.
 One should investigate to what extent the techniques used in this paper could
be modified
towards treating more general locally conformal transformations than projective
ones. This could turn out to be necessary
in order
to prove the  GGRT for the NSR-string along the lines of the GGRT
 laid out in \cite{LPP1,LPP2}. Such considerations should of course incorporate
also multi-string multi-loop vertices.
However, even if restricting ourselves to projective transformations it does
not seem entirely straightforward  to apply
 the techniques that have been developed here also to these more general
vertices.
This is related to the fact that these methods rely heavily on the
simplifications that occur
for the matrices $M$ etc when
particular  choices are made for the projective transformations used to
transport the basic dual vertices in a
four-string vertex.

We also wish to extend the methods developed here for Ramond four-string vertex
to four-string vertices involving
other twisted systems. Let us therefore recall the crucial steps of the Ramond
case discussed above which must carry over to these
other
cases. The first step was to find integral representations of the inverse
matrices $B^{ij}_{rs}$
 and the propagator $G(z,w)$ which factorize in the
following sense. For the latter one wants the $z$ and $w$ dependence to
factorize (apart from in the simple pole term).
This was however an immediate
 consequence of rewriting it as an integral; see \refbr{intrepofgg}.
In the former case  the factorization refers to the indices $r$ and $s$, and
for it to occur it seems necessary to make the particular
 choice \refbr{V1V2} of projective transformations (or any of the closely
related ones given in \cite{ENS2}).
 Namely, this choice implies that  $B^{ij}_{rs}$ may be written as in
\refbr{defofB} and that
$\*_{\l}M(\l)$ has rank zero\footnote{For the $V$'s used here this is easily
seen to true for any twisted system once the normal
ordering is
implemented by means of an untwisted normal ordering field as advocated in e.g.
\cite{ENS1}.},
 i.e. $\*_{\l}M(\l)\propto Rvv^T$ where $v$ is the infinite dimensional vector
associated with the massless oscillator modes.
 The whole four-fermion string vertex is therefore in this sense determined by
its massless sector. It is also of crucial
importance to have
$(1\pm M)^{-1}v$ etc expressed in terms of integral representations of
hypergeometric functions. All these
features are likely to generalize to other systems; one indication of this is
provided by the $\bf{Z}_3$ fermionic system
treated in \cite{EN1}.
Hence we believe that by means of these methods we can perform the crucial step
of ``resmoothing''
(i.e. deriving closed, geometrical expressions for)
arbitrary sewn twisted four-string vertices by means of pure operator sewing
methods.

\vspace{1 cm}
{\bf Acknowledgement}
\\We are very grateful to Christian Preitschopf for useful discussions.

\end{document}